\documentclass[showpacs,pre]{revtex4}
\usepackage{amssymb}
\usepackage{amsmath}
\usepackage{graphicx}
\usepackage{subfigure}

\begin{document}

\title{Gap solitons in quasiperiodic optical lattices}
\author{Hidetsugu Sakaguchi$^{1}$ and Boris A. Malomed$^{2}$}
\address{$^{1}$Department of Applied Science for Electronics and Materials,\\
Interdisciplinary Graduate School of Engineering Sciences,\\
Kyushu University, Kasuga, Fukuoka 816-8580, Japan\\
$^{2}$Department of Interdisciplinary Studies,\\
Faculty of Engineering, Tel Aviv University,\\
Tel Aviv 69978, Israel}

\begin{abstract}
Families of solitons in one- and two-dimensional (1D and 2D)
Gross-Pitaevskii equations with the repulsive nonlinearity and a potential
of the quasicrystallic type are constructed (in the 2D case, the potential
corresponds to a five-fold optical lattice). Stable 1D solitons in the weak
potential are explicitly found in three bandgaps. These solitons are mobile,
and they collide elastically. Many species of tightly bound 1D solitons are
found in the strong potential, both stable and unstable (unstable ones
transform themselves into asymmetric breathers). In the 2D model, families
of both fundamental and vortical solitons are found and are shown to be
stable.
\end{abstract}

\maketitle

\section{Introduction}

Solitons in Bose-Einstein condensates (BECs) are currently a subject of
intensive theoretical and experimental studies. Solitons supported by weak
attractive interactions between atoms were created in the condensate of $^{7}
$Li trapped in strongly elongated (nearly one-dimensional, 1D) traps \cite%
{Li} (although the actual shape of the solitons was actually nearly
three-dimensional, rather than nearly-1D; the latter feature was observed in
a salient form in the recent experiment \cite{Cornish}, where solitons were
created as a residual pattern after dynamical collapse took place in the $%
^{85}$Rb condensate). The stability of the solitons, with a given
number of atoms, against collapse is secured, in these settings,
by a combination of the trap's geometry and small absolute value
of the negative scattering length characterizing the attraction
between atoms ($\sim 0.1$ nm, in the case of $^{7}$Li, but up to
$\sim 2$ nm, in the condensate of $^{85}$Rb). A very accurate
description of the solitons is provided by the mean-field
approximation based on the Gross-Pitaevskii equation (GPE)
\cite{Pethik}.

Positive scattering length, which corresponds to repulsive interactions, is
more generic in BEC (in the above-mentioned experiments, the negative
scattering length was actually artificially induced by means of the Feshbach
resonance, both in $^{7}$Li and $^{85}$Rb). In a repulsive condensate,
\textit{gap solitons} (GSs) may be created as a result of the interplay of
the self-defocusing nonlinearity and periodic potential induced by an
optical lattice (OL, i.e., an interference pattern created by
counterpropagating laser beams illuminating the condensate) \cite{GS1D,GS}.
GSs emerge in \textit{bandgaps} of the system's linear spectrum, since the
combination of a \emph{negative} effective mass, appearing in a part of
bands adjacent to the gaps, with the repulsive interaction is exactly what
is needed to create a soliton. Theoretical models for GSs in BEC were
reviewed in Ref. \cite{Konotop}, and rigorous stability analysis for them in
the 1D case was developed in Refs. \cite{Pelinovsky}. Creation of a GS in
the experiment was reported in Ref. \cite{Oberthaler}, in $^{87}$Rb
condensate placed in a quasi-1D trap supplemented by a longitudinal OL; the
soliton contained a few hundred atoms. In a subsequent experiment,
large-size confined states with much larger numbers of atoms were discovered
in a stronger OL \cite{gap-wave-experiment}; an explanation to this
observation was recently proposed \cite{gap-wave-theory}, which,
essentially, treats the extended state as a segment of a nonlinear Bloch
wave, bounded by two fronts (domain walls) which are sustained by the strong
OL (a similar wall between filled and empty domains was predicted in BEC
with self-attraction in Ref. \cite{DW}).

Stable GSs were also predicted in 2D and 3D settings \cite{GS}, including 2D
solitons with embedded vorticity \cite{2Dvortex,2Dvortex2}. In nonlinear
optics, similar 2D spatial solitons \cite{PhotCryst} and 2D localized
vortices \cite{Ferrando} were predicted in photonic crystals and
photonic-crystal fibers, as well as in periodic photonic lattices induced by
perpendicular laser beams in photorefractive materials \cite{PRprediction}.
In media of the latter type, both fundamental \cite{Nature} and vortical 2D
solitons have been created in the experiment \cite{PRvortex}. A difference
from the self-repulsive BEC is the self-focusing character of the
nonlinearity in nonlinear optics (which is cubic in photonic crystals, and
saturating in photorefractive media); for this reason, the optical solitons
usually belong to the semi-infinite bandgap, where the effective mass is
always positive (although vortex solitons in a finite gap were also created
in a photorefractive medium \cite{Moti2gap}).

The GPE with self-attraction and periodic (optical-lattice) potential gives
rise to 2D solitons and vortices \cite{2Dvortex-attraction} similar to those
found in the above-mentioned optical models. Moreover, it has been recently
demonstrated that both the GPE with the self-focusing cubic term and OL
potential, and its counterpart with the saturable nonlinearity, that
appertains to photorefractives, give rise to stable higher-order localized
vortices (which are built as rings of unitary vortices) and \textit{%
supervortices} (similar rings with global vorticity imprinted onto them)
\cite{we}.

\textit{Quasiperiodic} OLs can be easily created too: in the 1D case, as a
superposition of two sublattices with incommensurate periods, and in the 2D
case, as a combination of $N=5$ or $N\geq 7$ quasi-1D sublattices with wave
vectors $\mathbf{k}^{(n)}$ of equal lengths, which make equal angles $2\pi /N
$ between themselves. In particular, the 2D lattice with $N=5$ is known as
the \textit{\ Penrose tiling }(PT). The bandgap spectrum of 2D photonic
crystals of the PT type has been studied in detail \cite{Penrose}, with a
conclusion that true (omnidirectional) bandgaps may be supported by the PT.

The use of quasiperiodic lattices, in one and two dimensions
alike, offers new degrees of freedom that allow one to
\textit{engineer} desirable bandgaps in the spectrum and, this
way, \emph{design} soliton families in nonlinear media. Several
theoretical works sought for solitons in models combining
quasiperiodic lattice potentials and cubic \emph{self-focusing}.
In an early work \cite{no-soliton}, solitons were not found in a
1D quasiperiodic model; however, they were later discovered in a
``deterministic aperiodic" discrete nonlinear Schr\"{o}dinger
(NLS) equation, which may be regarded as a limit case of the 1D
model with a very strong quasiperiodic potential \cite{discrete}.
Localized and delocalized solutions were also studied in the 1D
model with a superlattice (i.e., an ordinary OL subjected to an
additional modulation, which is different from a quasicrystal)
\cite{superlattice}. Recently, 2D solitons (with zero vorticity)
were numerically constructed in a model of a photonic crystal made
of a self-focusing material, with $N=12$, in terms of the above
definition \cite{Xie}.

The objective of this work is to find GSs supported by the interplay of the
cubic \emph{repulsive} nonlinearity and quasiperiodic lattice potentials. We
will report systematic results for fundamental solitons in the 1D and 2D
models (the latter one will be elaborated for the PT lattice). In the 1D
case, loosely bound GSs in the weak lattice are mobile. Tightly bound GSs in
the strong lattice are found in many modifications, some stable and some
not. Stable vortex solitons in the PT potential will be demonstrated too.

\section{One-dimensional solitons}

In the mean-field approximation, the evolution of the single-atom function $%
\phi $ obeys the GPE with the repulsive nonlinear term and quasiperiodic
potential $U(x)$. In the normalized form, the equation is
\begin{equation}
i\frac{\partial \phi }{\partial t}=-\frac{1}{2}\frac{\partial ^{2}\phi }{
\partial x^{2}}+|\phi |^{2}\phi +U(x)\phi ,  \label{GPE}
\end{equation}
As said above, the 1D potential is a combination of two incommensurate
spatial harmonics with equal amplitudes $\varepsilon $,
\begin{equation}
U(x)=-\varepsilon \lbrack \cos \left( \pi (x-L/2)\right) +\cos \left( q\pi
(x-L/2)\right) ],  \label{U1D}
\end{equation}
where $x=L/2$ is the central point of a large($L\gg 1$) trapping domain, $%
0<x<L$, the period of the first sublattice is normalized to be $2$, and $q$
is an irrational number. Below, we display results for $q=(\sqrt{5}
+1)/2\approx \allowbreak 1.62$.

As is known, an exact bandgap spectrum of the linear Schr\"{o}dinger
equation with a quasi-periodic potential is fractal. Without the aim to
display the spectrum in full detail, in Fig. 1 we show its part which is
relevant to the quest for gap solitons. Bands of values of chemical
potential $\mu $ corresponding to families of quasi-Bloch states,
\begin{equation}
\phi (x,t)=e^{-i\mu t}\chi (x),  \label{varphi}
\end{equation}%
are covered by vertical segments at fixed values of OL strength $\varepsilon
$. In panel 1(a), one can discern four finite gaps (discontinuities between
the bands, the lowest gap being extremely narrow) and the underlying
semi-infinite gap (one extending to $\mu \rightarrow -\infty $). A result of
the standard perturbation theory is that these gaps start (at $\varepsilon =0
$) at $\mu =\pi ^{2}/2$ and $\mu =(\pi q)^{2}/2$. Panel 1(b) details the
results in a narrower range of $\mu $ but for an extended interval of $%
\varepsilon $. One can observe, in particular, that the lowest band in
Fig.~1(a) splits into two bands at $\varepsilon =0.8$, and three bands at $%
\varepsilon =1$ in Fig. 1(b). With the increase of $\varepsilon $,
additional bandgaps open up on a finer scale (accurate computation of the
spectrum becomes rather difficult for $\varepsilon >2$).
\begin{figure}[tbp]
\includegraphics[height=4.cm]{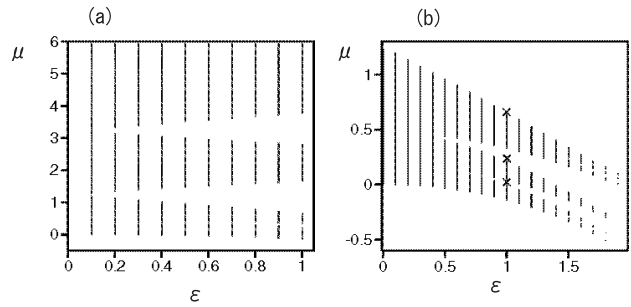}
\caption{A part of the bandgap structure in the linear version of Eq. (%
\protect\ref{GPE}). Crosses in (b) indicate values of $\protect\mu $ for
three GSs shown in Fig.~2. }
\label{fig1}
\end{figure}

As the effective mass is negative near the top of (quasi-)Bloch bands, GSs
are expected to exist in adjacent lower parts of the gaps. Assuming real $%
\chi (x)$ and varying $\mu $ in Eq. (\ref{varphi}), we numerically searched
for a family of solutions of the corresponding stationary version of Eq. (%
\ref{GPE}), $\chi ^{\prime \prime }/2=\chi ^{3}+U(x)\chi -\mu \chi
$, satisfying boundary conditions $\chi (x=L/2)=A$, $\chi ^{\prime
}(x=L/2)=\chi (x=L)=0$. Figure 2 displays three typical examples
of GSs with equal amplitudes, $\chi (x=L/2)=A=0.3$, that were
found in three finite bandgaps for $\varepsilon =1$ at values of
$\mu $ marked by crosses in Fig. 1(b) [as usual (in the GPE with
self-repulsion), no solitons in the are found in the semi-infinite
gap]. A blow-up of the GS in Fig.~2(b) is shown in Fig.~2(d), to
demonstrate that, while the envelope of the solution is that of a
solitary wave, the carrier wave function is nearly quasiperiodic.
\begin{figure}[tbp]
\includegraphics[height=6.cm]{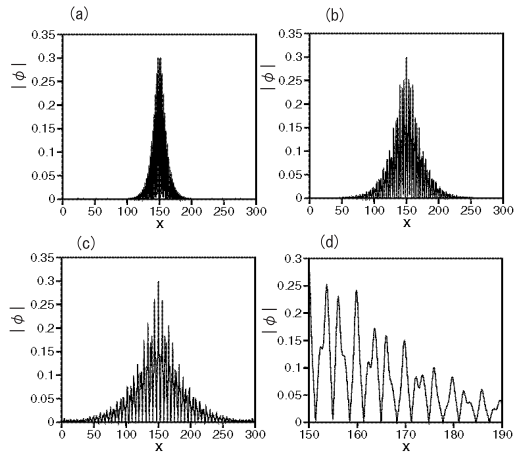}
\caption{Examples of stable gap solitons in the 1D model for $\protect%
\varepsilon =1$, with the chemical potential and norm $\protect\mu =0.658$, $%
N=0.104$ (a), $\protect\mu =0.237$,\thinspace $N=0.763$ (b), and $\protect%
\mu =0.0196$,\thinspace $N=0.935$ (c). Panel (d) is a blow-up of (b) in the
region of $150<x<190$.}
\label{fig2}
\end{figure}

Families of the GS solutions are characterized by the dependence of $\mu $
on the soliton's norm, $N=\int_{-\infty }^{+\infty }\chi ^{2}(x)dx$. For GS
families in the three lowest finite bandgaps, whose examples were displayed
in Fig. 2, the $N(\,\mu )$ dependences are plotted in Fig. 3. Naturally, $%
N\rightarrow 0$ corresponds to $\mu $ approaching the border of
the quasi-Bloch band located under the respective gap.

The numerically found solution branches do not cover the range of
$\mu$ corresponding to the entire bandgap in each case. However,
the effective termination of the branches inside the gaps seems to
be a numerical problem, rather than a true feature of the model:
it is difficult to secure convergence of the numerical solutions
for very large values of $N$.

Stability analysis of the GSs was performed by means of direct
simulations, using the split-step method with 4096 Fourier modes.
This way, it was concluded that all the three families displayed
in Fig. 3 are completely stable.
\begin{figure}[tbp]
\includegraphics[height=3.5cm]{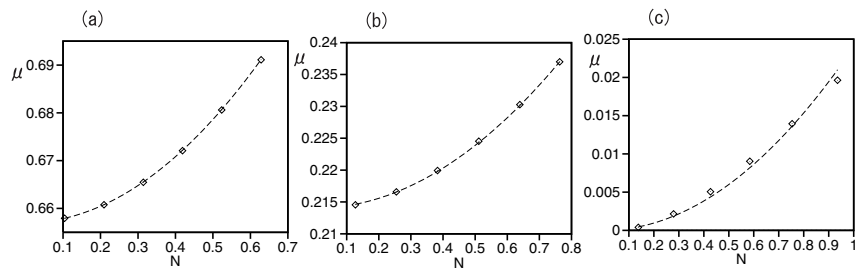}
\caption{The chemical potential vs. norm for gap-soliton families in three
lowest finite bandgaps (a,b,c) of the 1D model with $\protect\varepsilon =1$%
. The dashed curves are added as guides to the eye.}
\label{fig3}
\end{figure}

GSs obtained above can be readily set in motion the same way as it
was done in Ref. \cite{we-earlier}, i.e., multiplying a stationary
solution by $\exp (ikx)$, with $k$ not too large. Examples are
displayed in Fig. 4, where the ``shove factor" and the resulting
average velocity of the
moving soliton are $k=0.03$, $v=-0.104$ (a); $k=0.015$, $v=-0.117$ (b); and $%
k=0.005$, $v=-0.03$ (c). Accordingly, the effective GS mass $M\equiv k/v$ is
$-0.288$ (a), $-0.128$ (b), and $-0.17$ (c). The mass is negative, as it
should be for GSs \cite{we-earlier}. On the other hand, moving solitons
feature conspicuous radiation losses if they are set in motion by shove $k$
exceeding a certain critical value, which is $k_{\mathrm{cr}}\approx 0.05$, $%
k\approx 0.017$, and $k\approx 0.008$ for the solitons from Figs. 2(a),
2(b), and 2(c) respectively, although the transition to the lossy motion
regime is not very sharp.
\begin{figure}[tbp]
\includegraphics[height=3.5cm]{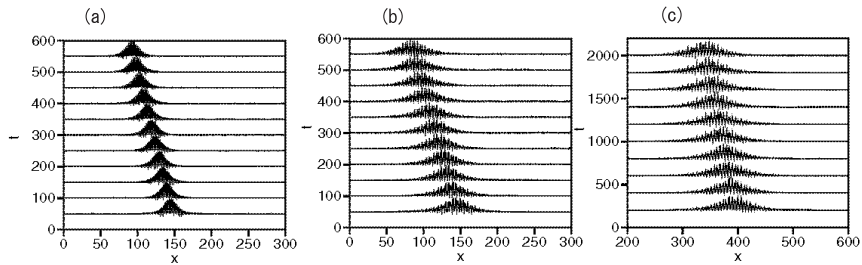}
\caption{Examples of stable moving gap solitons in three finite bandgaps
(a,b,c) of the 1D model.}
\label{fig4}
\end{figure}

Following the approach based on the separation of rapidly and slowly varying
functions, one can approximate GS solutions by
\begin{equation}
\phi (x,t)=e^{-i\mu t}\varphi _{\mu }(x)\Phi (x,t),  \label{Phi}
\end{equation}
where $\varphi _{\mu }(x)$ is a linear quasi-Bloch function pertaining to
chemical potential $\mu $, and $\Phi (x,t)$ is a slowly varying amplitude,
for which an averaged NLS equation can be derived as in Ref. \cite%
{we-earlier},
\begin{equation}
i\frac{\partial \Phi }{\partial t}=-\frac{1}{2M}\frac{\partial ^{2}\Phi }{%
\partial x^{2}}+G|\Phi |^{2}\Phi .  \label{aver}
\end{equation}%
Here $M<0$ is the above-mentioned effective mass, and $G>0$ is an effective
nonlinearity coefficient, which can be found by matching obvious soliton
solutions to Eq. (\ref{aver}), $\Phi =A\mathrm{sech}\left( A\sqrt{-MG}%
x\right) $ ($A$ is an arbitrary amplitude), to the envelope of the
numerically found GSs. As a result, for the three quiescent GSs displayed in
Fig.~2 we find $G=0.68$ (a), $G=0.41$ (b), and $G=0.13$ (c). The
availability of stable moving solitons suggests to consider collisions
between them. As predicted by the averaged equation (\ref{aver}), moving GSs
emerge unscathed from collisions, see an example in Fig. 5.
\begin{figure}[tbp]
\includegraphics[height=4.5cm]{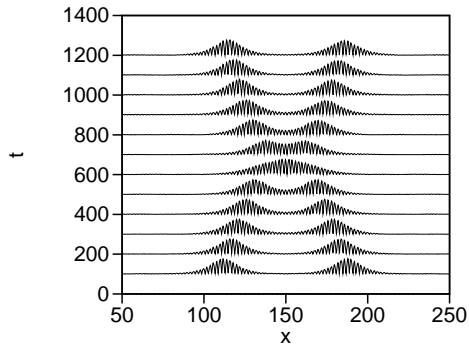}
\caption{Head-on collision of two gap solitons, which are obtained from ones
displayed in Fig. 2(a) by shoving them with $\exp \left( \pm ikx\right) $, $%
k=0.02$.}
\label{fig5}
\end{figure}

The GSs displayed above are \textit{loosely bound} solutions, in terms of
Ref. \cite{we-earlier}, as they are generated by the GPE with a relatively
small strength ($\varepsilon $) of the pinning potential. For larger $%
\varepsilon $, bands between the gaps in the model's spectrum become nearly
invisible, and the character of GS solutions drastically alters, with a
large variety of \textit{tightly bound} solitons found in the broad
bandgaps. Figure 6(a) displays four different kinds of the solutions (for $%
\varepsilon =5$) with equal amplitudes, $|\phi (x=L/2)|=2$. In the first
potential well around the central point, all the profiles are almost
identical, but the next pair of peaks appear at markedly different
positions, and may even have different signs. The form of the quasiperiodic
potential $U(x)$ is displayed below the GSs. Naturally, local maxima of $%
|\phi |$ are found near minima of the potential. In fact, more GS species
can be found, in addition to the four ones shown here. Figure 6(b)
quantifies four GS families, whose representatives are displayed in Fig.
6(a), in terms of the $\mu (N)$ dependence. The curves overlap at
sufficiently small $N$, as in this limit each GS species virtually reduces
to the single peak in the first potential well; however, the species become
very different at larger $N$, corresponding to the different GS shapes
displayed in Fig. 6(a). For example, branch 4 starts to deviate from the
other three ones near $N=0.5$, when negative peaks begin to develop near $%
x=16$ and $24$.
\begin{figure}[tbp]
\includegraphics[height=4.5cm]{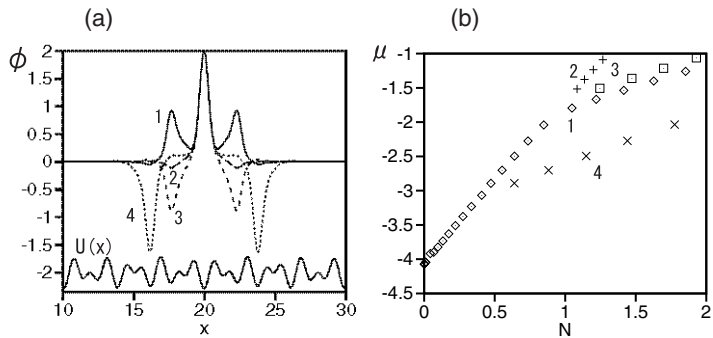}
\caption{(a) Four types of stationary gap solitons found in the 1D model
with a strong optical lattice, $\protect\varepsilon =5$. The chemical
potential $\protect\mu $ and the norm $N$ are, respectively, $\protect\mu %
=-1.261,N=1.853$ (1), $\protect\mu =-1.088,N=1.266$ (2), $\protect\mu %
=-1.214,N=1.697$ (3), $\protect\mu =-1.2415,N=2.854$ (4). The form
of the potential is shown in the bottom. Panel (b) displays curves
$\protect\mu (N)$ for the four species of the gap solitons.
Solution branches shown in (b) terminate at points where the
solitons quickly develop the tail structures.} \label{fig6}
\end{figure}

Direct simulations demonstrate that the GS species labeled as 1, 2 and 4 in
Fig. 6 are \emph{stable}, while the family labeled by 3 is \emph{unstable}.
Further, Fig. 7(a) displays a typical example of the evolution of an
unstable GS. The instability breaks the soliton's symmetry and makes it a
breather. Although the breathers's amplitude features complex oscillations,
see Fig. 7(b), the breather does not decay, maintaining its localized shape.
\begin{figure}[tbp]
\includegraphics[height=5.cm]{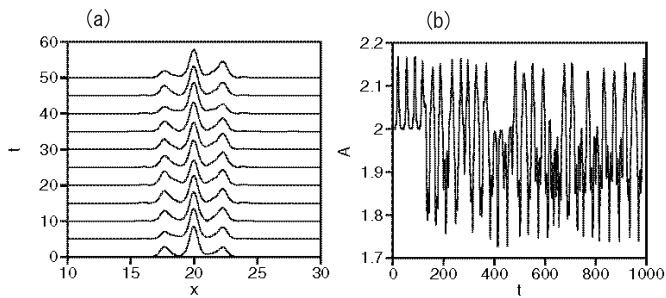}
\caption{(a) Evolution of $|\protect\phi |$ for an unstable gap soliton
labeled by 3 in Fig. 6(a). (b) Evolution of $|\protect\phi (L/2)|$ in the
gap soliton. }
\label{fig7}
\end{figure}

\section{Two-dimensional solitons}

In the normalized form, the 2D version of the GPE with the PT (Penrose
tiling) potential is
\begin{equation}
i\frac{\partial \phi }{\partial t}=-\frac{1}{2}\nabla ^{2}\phi +|\phi
|^{2}\phi -\varepsilon \left[ \sum_{n=1}^{5}\cos \left( \mathbf{k}%
^{(n)}\cdot \mathbf{r}\right) \right] \phi ,  \label{GPE2D}
\end{equation}%
with $\left\{ k_{x}^{(n)},k_{y}^{(n)}\right\} =\pi \left\{ \cos \left( 2\pi
(n-1)/5\right) ,\sin \left( 2\pi (n-1)/5\right) \right\} $. GS solutions in
two dimensions were constructed by means of a combination of the known
method of the integration in imaginary time \cite{imaginary} and subsequent
real-time simulations. To this end, a substitution was first made, $\phi
(x,y,t)=e^{-i\mu t}\Phi (x,y,-i\tau )$, which transforms Eq. (\ref{GPE2D})
into an nonlinear diffusion equation,
\begin{equation}
\frac{\partial \Phi }{\partial \tau }=\frac{1}{2}\nabla ^{2}\Phi +(\mu
_{0}-\Phi ^{2})\Phi +\varepsilon \left[ \sum_{n=1}^{5}\cos \left( \mathbf{k}%
^{(n)}\cdot \mathbf{r}\right) \right] \Phi .  \label{IMGPE}
\end{equation}%
This equation was solved numerically by dint of the split-step Fourier
method with $256\times 256$ modes, in a domain of the size of $50\times 50$.
It was observed that the norm of the solution originally decreased, and then
began to increase. The imaginary-time integration was switched back into
simulation of the GPE in real time when the norm attained its minimum.

First, we present typical examples of stable fundamental (zero-vorticity) 2D
solitons, for $\mu =-1.56$, $-3.33$, and $-0.41$. They were generated by
initial conditions [in Eq. (\ref{IMGPE})] with, respectively,
\begin{equation}
\Phi _{0}(x,y)=\left\{
\begin{array}{c}
\exp \left( -0.5(x^{2}+y^{2})\right)  \\
1.5\exp \left( -0.15(x^{2}+y^{2})\right)  \\
2\exp \left( -0.05(x^{2}+y^{2})\right)
\end{array}%
\right\} \sum_{n=1}^{5}\cos \left( \frac{1}{2}\mathbf{k}^{(n)}\cdot \mathbf{r%
}\right) ,  \label{Phi0}
\end{equation}%
the last multiplier being a half-harmonic counterpart of the PT potential in
Eq. (\ref{GPE2D}), which is expected to parametrically couple to the
potential.

Figure 8 displays contour plots of $|\phi (x,y)|$ in final states produced
by the numerical integration. To illustrate the proximity of the solutions
to truly stationary ones, and their stability, in Fig. 9 we show the
asymptotic time dependence of the peak amplitudes of the three solitons at $%
(x,y)=(L/2,L/2)=(25,25)$. These gap solitons are found in a strong
optical lattice with $\varepsilon =5$, for which bands separating
the spectral gaps are extremely narrow.
\begin{figure}[tbp]
\includegraphics[height=5.cm]{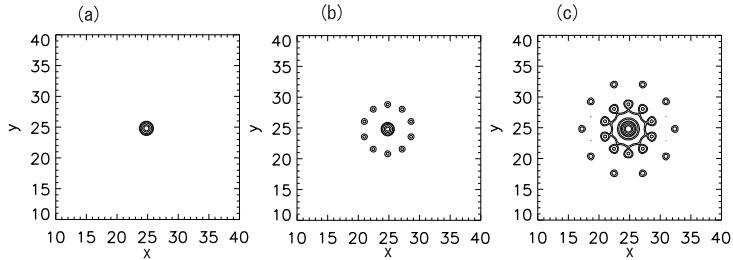}
\caption{Contour plots of $|\protect\phi (x,y)|$ for three examples of
stable gap solitons in the 2D model with a strong Penrose-tiling optical
lattice, $\protect\varepsilon =5$. The solitons are generated by initial
conditions (\protect\ref{Phi0}), with, respectively, $\protect\mu =-1.56$
(a), $\protect\mu =-3.33$ (b), and $\protect\mu =-0.41$ (c). The norms of
the three solitons are $N_{a}=147$, $N_{b}=103$, and $N_{c}=268$.}
\label{fig8}
\end{figure}
\begin{figure}[tbp]
\includegraphics[height=3.5cm]{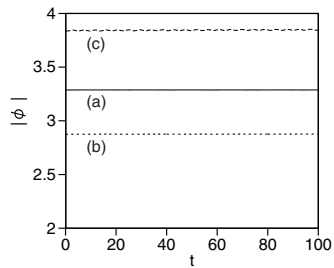}
\caption{Time dependence of the peak amplitudes of three localized solutions
shown in Fig.~8.}
\label{fig9}
\end{figure}

Systematic results for fundamental 2D solitons were generated by using
initial conditions (\ref{Phi0}) and varying the respective chemical
potential $\mu $. Figure 10 shows $\mu (N)$ dependences for three families
of thus generated GSs, where $N=\int \int \left\vert \phi (x,y)\right\vert
^{2}dxdy$ is the usual 2D norm. At relatively small values of $N$, the
dependences completely overlap, and they slightly split up at very large $N$%
, corresponding to $\mu \rightarrow -0$. They are somewhat (but not quite)
similar to plots for tightly bound GSs in the 1D model, cf. Fig.~6(b). There
are no loosely bound stable GSs in two dimensions, because the averaged
two-dimensional NLS equation [a 2D counterpart of Eq. (\ref{aver})] does not
have stable soliton solutions.
\begin{figure}[tbp]
\includegraphics[height=3.5cm]{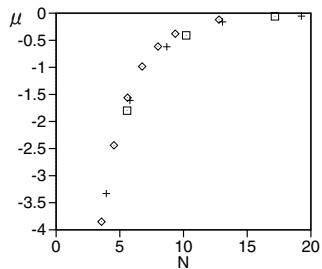}
\caption{Chemical potential $\protect\mu $ of two-dimensional fundamental
solitons vs. $N$, in the model with $\protect\varepsilon =5$. Rhombuses,
crosses, and squares represent soliton families generated, respectively, by
three initial configurations in Eq. (\protect\ref{Phi0}) (typical examples
of solitons belonging to the three families are given in Fig. 8).}
\label{fig10}
\end{figure}

Lastly, stable 2D solitons with \emph{embedded vorticity} $S$ can
be found too. They were generated by adding a factor, $\left(
x^{2}+y^{2}\right) ^{|S|/2}\exp (iS\theta )$, which distinguishes
vortex states in uniform media, to initial \textit{ans\"{a}tze}
(\ref{Phi0}). A typical example of such a stable \textit{vortical
gap soliton} in presented Fig. 11. While the field pattern of
$\left\vert \phi (x,y)\right\vert $, displayed in panel
11(a), is stationary, panel 11(b) is actually a snapshot, as the pattern of $%
\mathrm{\ Re~}\phi (x,y)$ shown in this panel rotates clockwise, at the
angular velocity of $\omega =\mu /S=-3$.
\begin{figure}[tbp]
\includegraphics[height=3.5cm]{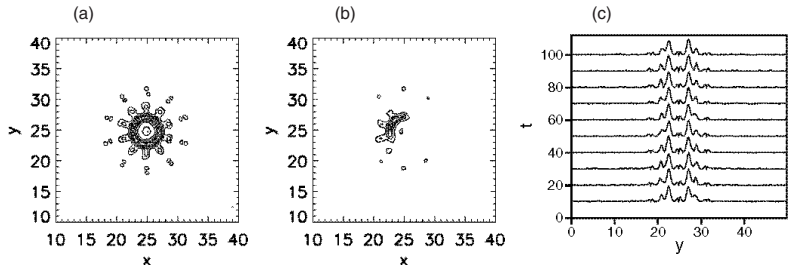}
\caption{A typical example of a stable vortical gap soliton with $S=-1$,
found in the model with $\protect\varepsilon =5$. (a) and (b): Contour plots
of $\left\vert \protect\phi (x,y)\right\vert $ and $\mathrm{Re~}\protect\phi %
(x,y)$, the latter shown only in the region with $\mathrm{Re~}\protect\phi %
(x,y)>0$. (c): A sequence of central cross-sections, $\left\vert \protect%
\phi (x=L/2,y)\right\vert $, taken at different moments of time, which
demonstrate the vanishing of the field at the center, as must be in the
vortex, and its stability. The norm and chemical potential of the solution
are $N=254$ and $\protect\mu =-3$.}
\label{fig11}
\end{figure}

\section{Conclusion}

We have constructed families of 1D and 2D gap solitons in the model
combining the self-defocusing nonlinearity and quasiperiodic lattice
potentials. Loosely bound 1D solitons supported by a weak lattice were found
in three gaps; they are mobile, and collide elastically. The strong 1D
lattice supports a variety of species of tightly bound solitons. Most of
them are stable, unstable ones giving rise to robust asymmetric breathers.
2D gap solitons, both fundamental and vortical ones, are stable too, and
they may only have a tightly-bound shape. The predicted solitons can be
created by means of available experimental techniques in a Bose-Einstein
condensate with a positive scattering length, loading it into an optical
lattice induced by a superposition of several laser beams.

B.A.M. appreciates hospitality of the Department of Applied Science for
Electronics and Materials at the Interdisciplinary Graduate School of
Engineering Sciences, Kyushu University (Fukuoka, Japan). This work was
supported, in a part, by the Grant-in-Aid for Scientific Research
No.17540358 from the Ministry of Education, Culture, Sports, Science and
Technology of Japan, and the Israel Science Foundation through a
Center-Excellence grant No. 8006/03.

\end{document}